# Conversion of the chemical concentration of odorous mixtures into odour concentration and odour intensity: a comparison of methods

Chuandong Wu [1], Jiemin Liu [1], Peng Zhao [2], Martin Piringer [3], Günther Schauberger [4*]

[1] University of Science and Technology Beijing, School of Chemistry and Biological Engineering, Beijing, China, en.ustb.edu.cn
[2] Beijing Municipal Institute of Labor Protection, Key Laboratory of Occupational Health and Safety, Beijing, China, www.bmilp.com
[3] Department of Environmental Meteorology, Central Institute of Meteorology and Geodynamics, Vienna, Austria; www.zamg.ac.at
[4] WG Environmental Health, Unit for Physiology and Biophysics, University of Veterinary Medicine, Vienna, Austria, www.vetmeduni.ac.at, *corresponding author

**Address**
* Corresponding author:

Assoc. Prof.
Dr. Günther Schauberger
University of Veterinary Medicine
Department of Biomedical Sciences
WG Environmental Health
Veterinärplatz 1
A 1210 Vienna
Austria
++ 43 1 25077 4574
gunther.schauberger@vetmeduni.ac.at

**Research Highlights**

- Concentration of odorous substances as surrogate for odour concentration/intensity
- Comparison of conversion methods with various degrees of complexity
- Model evaluation by seven VOCs: 23 binary mixtures and 5 mixtures of 7 substances
- Model input: odour threshold concentration and the slope of the Weber-Fechner law
- No further calibration by olfactometric measurements necessary

# Abstract

Continuous odour measurements both of emissions as well as ambient concentrations are seldom realised, mainly because of their high costs. They are therefore often substituted by concentration measurements of odorous substances. Then a conversion of the chemical concentrations $C$ (mg m$^{-3}$) into odour concentrations $C_{OD}$ (ou$_E$ m$^{-3}$) and odour intensities $OI$ is necessary. Four methods to convert the concentrations of single substances to the odour concentrations and odour intensities of an odorous mixture are investigated: (1) direct use of measured concentrations, (2) the sum of the odour activity value $SOAV$, (3) the sum of the odour intensities $SOI$, and (4) the equivalent odour concentration $EOC$, as a new method. The methods are evaluated with olfactometric measurements of seven substances as well as their mixtures. The results indicate that the $SOI$ and $EOC$ conversion methods deliver reliable values. These methods use not only the odour threshold concentration but also the slope of the Weber-Fechner law to include the sensitivity of the odour perception of the individual substances. They fulfil the criteria of an objective conversion without the need of a further calibration by additional olfactometric measurements.

## 1. Introduction

In the field of environmental odour, it is difficult to realise continuous odour measurements of emission as well as ambient concentrations in the vicinity of an odour source. In many cases, odour measurements are substituted by concentration measurements of odorous substances. This indirect method has several reasons. (1) olfactometric measurements need air sampling and several panellists for the measurement, therefore the costs are high, (2) the measurements can only be done discontinuously, and usually inside an odourless laboratory, and (3) in many cases only the emission concentration can be measured, because ambient odour concentrations are often too low to get reliable results (Gostelow et al., 2003).

Continuous odour measurements would, however, be desirable as they can be seen as a prerequisite for several applications. Emission concentration measurements would be required for dispersion modelling to assess the ambient concentration and the related odour annoyance. Measurements of the ambient concentration of odorous substances could be used by environmental protection agencies to monitor the odour annoyance caused by a plant at a certain site (e.g. Kabir and Kim (2010), Schauberger et al. (2011)), and from ambient concentrations of odorous substances, the emission flow rate could be back-calculated by inverse modelling (e.g. Schauberger et al. (2013); Schauberger et al. (2008)). Moreover, it is agreed that the use of chemical concentrations instead of odour measurements should be limited to cases where the odour concentration is – for some reason – not directly measurable.

Using concentration measurements of odorous substances instead of olfactometric measurements, a conversion of the chemical concentrations $C$ (mg m$^{-3}$) into odour concentrations $C_{OD}$ (ou$_E$ m$^{-3}$) and odour intensities $OI$ is necessary, for which several concepts are in use. The simplest approach is the direct use of the concentration $C$ of a single substance. The sum of the concentration values is then used as a surrogate for the measured odour concentration $C_{OD}$. This method works well for single substances (e.g. H$_2$S (Dincer and Muezzinoglu, 2007; Gostelow et al., 2001)) and for a group of substances with a constant composition. To determine the parameters of a regression between concentration and odour concentration, olfactometric measurements have to be performed.

The second concept, called odour activity value $OAV$, is based on the normalisation of the concentration $C$ by the odour threshold concentration $C_{OT}$. If more than one substance is involved in the odour perception, then the sum of the individual $OAV$ is used. This value is called sum of the odour activity value $SOAV$ (Capelli et al., 2013; Parker et al., 2012).

A more sophisticated conversion is using not only the odour threshold concentrations of individual substances but also the slope $k$ of the odour intensity - odour concentration relationship (Kim and Park, 2008). It is predominantly used for air quality assessments in Korea. This concept is based on the idea that the perception of an odorous mixture can be calculated by the sum of the individual odour intensities $SOI$ and the related odour concentrations $C_{OD}$ (Kim, 2010; Kim and Kim, 2014b).

The new concept of the equivalent odour concentration $EOC$ introduced here uses the physiological rule that the perception of odour intensity can be assessed by the sum of the stimuli which can be determined by the odour concentration of the individual substances, taking into account the sensitivity of the perception by the slope of the Weber-Fechner law. The goal of the conversion is an objective method, which can be used without the need of an additional calibration by olfactometric measurements.

The conversion from the chemical concentration of single substances to the odour concentrations and odour intensities of an odorous mixture using the four methods is the central topic of this paper. The ability of the four conversion methods to produce reliable odour concentrations is investigated here by comparing them with olfactometric odour concentration measurements; also the odour intensities will be compared. The comparisons will be undertaken both for the single substances as well as their mixtures.

## 2. Materials and Methods

### 2.1 Conversion methods

The conversion of the concentration measurements of individual substances $C_i$ to odour concentrations of the mixture $C_{OD}$ and odour intensity $OI$ is done by the four different methods briefly outlined in the introduction; the equations used are summarised in Tab. 1. Besides the concentration of each substance $C_i$, which is used by all four methods, the necessary additional input is given.

Tab. 1: Formula apparatus for the conversion of the concentration of single substances $C_i$ (mg m$^{-3}$) into odour concentrations $C_{OD}$ (ou$_E$ m$^{-3}$) and odour intensities $OI$ (-) of an odorous mixture by four different methods: (1) method based on the concentration $C$, (2) the sum of the odour activity values $SOAV$, (3) the sum of odour intensities $SOI$, and (4) the equivalent odour concentration $EOC$. Additional input is the odour threshold concentration $C_{OT,i}$ (mg m$^{-3}$) and/or the slope $k_i$ of the Weber-Fechner law. The parameters of the reference substance Ethyl acetate are denoted by $j$.

| Conversion method | Additional input | Odour concentration $C_{OD}$ (ou$_E$/m³) | Odour Intensity $OI$ (-) |
|---|---|---|---|
| $C$ | - | $C_{OD}{}^C = k_c \sum C_i / m_{OD,0}$ | $OI^C = \log C_{OD}{}^C + 0.5$ |
| $SOAV$ | $C_{OT,i}$ | $SOAV = \sum C_i / m_{OD,i}$ | $OI^{SOAV} = \log SOAV + 0.5$ |
| $SOI$ | $C_{OT,i}, k_i$ | $C_{OD}{}^{SOI} = 10^{\frac{SOI - 0.5}{k_j}}$ | $SOI = \log \sum 10^{OI_i}$ <br> $OI_i = k_i \log {C_i}/{m_{OD,i}} + 0.5$ |
| $EOC$ | $C_{OT,i}, k_i$ | $EOC_j = \sum_{i=1}^{n} 10^{\frac{k_i}{k_j} \log \frac{C_i}{m_{OD,i}}}$ | $OI^{EOC_j} = k_j \log EOC_j + 0.5$ |

$m_{OD,i} = C_{OT,i} / C_{OD,0}$ with $C_{OD,0}$ = 1 ou$_E$ m$^{-3}$, $k_c$ is the proportionality constant

The first method uses the measured concentrations without any further information to assess the odour concentration according to $C_{OD}{}^C = k_c \sum C_i / m_{OD,0}$, by using the specific odour mass set to unity $m_{OD,0}$ = 1 mg ou$^{-1}$ to reach a proper measuring unit of the odour concentration (ou$_E$ m$^{-3}$). The proportionality constant $k_c$ can be determined by olfactometric measurements and a linear regression analysis (e.g. Dincer and Muezzinoglu (2007)). In some cases also non-linear functions are in use to describe the relationship between concentration and odour concentration (e.g. power function for H$_2$S (Franke et al., 2009; Gostelow et al., 2001)). The $OI$ is calculated from the odour concentration by using the Weber-Fechner law (see in detail Section 2.4) with an assumed slope of $k$ = 1.0 which results in $OI^C = \log C_{OD}{}^C + 0.5$.

The second conversion method uses the odour threshold concentration $C_{OT,i}$ of each odorous substance to calculate the odour activity value $OAV_i$ of a certain chemical substance $C_i$ (mg m$^{-3}$), calculated by $OAV_i = C_i / m_{OD,i}$, using the specific odour mass $m_{OD,i}$, to get the proper measuring unit of an odour concentration (ou$_E$ m$^{-3}$). The $OAV$ of the entire mixture $SOAV$ is then calculated by the sum of the individual odour activity values $SOAV = \sum OAV_i$ which correspond to an odour concentration. The odour intensity of this mixture is then calculated by the Weber-Fechner law under the assumption of a unity slope $k = 1$ by $OI^{SOAV} = \log SOAV + 0.5$.

The third method was proposed by Kim and Park (2008) using the odour threshold concentration $C_{OT,i}$ (respectively the derived specific odour mass $m_{OD,i}$) and the odour intensity $OI_i$ calculated by the Weber-Fechner law for each single substance. The sum of the odour intensities $SOI$ of the entire mixture is then calculated by $SOI = \sum 10^{OI_i}$. The backward calculation of the odour intensity $C_{OD}^{SOI}$ is done for a selected substance $j$ by the Weber-Fechner law (Kim, 2010).

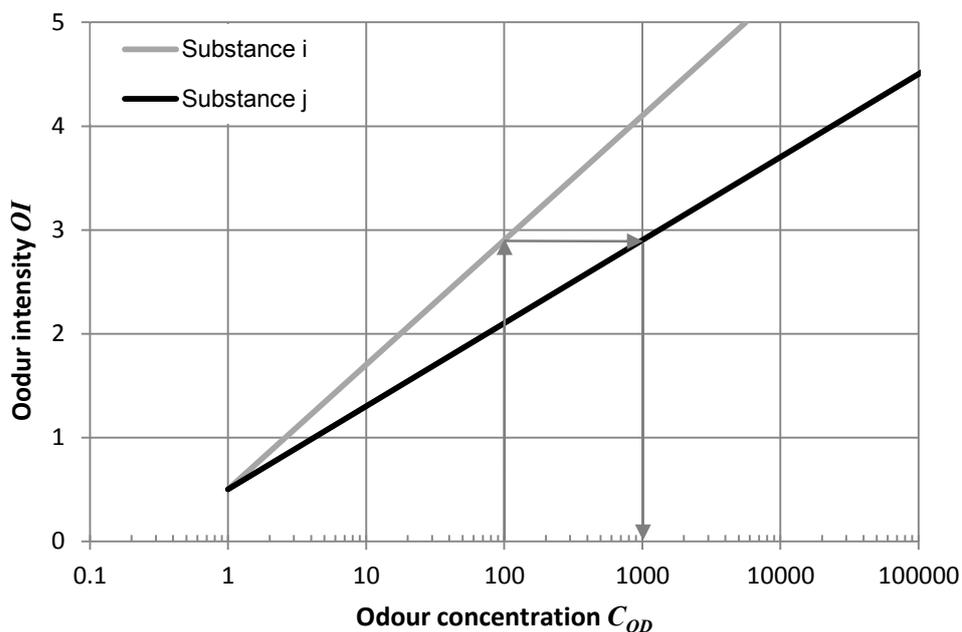

Fig. 1: Schematic representation of the conversion by the equivalent odour concentration $EOC$ method for two odorous substances $i$ and $j$ (reference substance with the shallow slope) with 100 ou$_E$ m$^{-3}$. The equivalent odour concentration of substance $i$ yields the same odour intensity which results in 1000 ou$_E$ m$^{-3}$.

The conversion method introduced here, called equivalent odour concentration $EOC$, is based on the odour threshold concentration $C_{OT,i}$ and the slope of the Weber-Fechner law $k_i$. The equivalent odour concentration $EOC_j$ related to one selected substance $j$ of the mixture can be calculated according to $EOC_j = \sum 10^{\frac{k_i}{k_j} \log C_{OD,i}}$ which corresponds to an odour concentration. The odour intensity of this odorous mixture is then calculated by the Weber-Fechner law with $OI^{EOC_j} = k_j \log EOC_j + 0.5$. The $EOC_j$ of a selected substance $j$, represents the odour concentration of the selected substance $j$ which is necessary to perceive the odour concentration of the entire mixture of substances. The conversion of a binary mixture of two odorous substances is depicted in Fig. 1. The odour concentration of substance $i$ is $C_{OD,i}$ = 100 ou$_E$ m$^{-3}$, for substance $j$, $C_{OD,j}$ = 100 ou$_E$ m$^{-3}$. The slopes are assumed with $k_i$ = 1.2 and $k_j$ = 0.8, respectively. The conversion of the odour concentration of substance $i$ to the reference substance $j$ with the smaller slope is shown by grey arrows. The equivalent odour concentration related to the reference substance $j$ will result in $EOC_j$ = 100 ou$_E$ m$^{-3}$ + 1000 ou$_E$ m$^{-3}$ = 1100 ou$_E$ m$^{-3}$.

## 2.2 Chemical substances

Odour concentration measurements of seven chemical substances often emitted by the petrochemical industry were available for this investigation. The chemical characteristics of these substances are presented in Tab. 2.

Tab. 2: Chemical characteristics of the odorous monomolecular substances used in the present investigation.

| Substance | Butyl acetate | Benzene | Ethyl acetate | Toluene | m-Xylene | o-Xylene | α-Pinene |
|---|---|---|---|---|---|---|---|
| | $C_6H_{12}O_2$ | $C_6H_6$ | $C_4H_8O_2$ | $C_7H_8$ | $C_8H_{10}$ | $C_8H_{10}$ | $C_{10}H_{16}$ |
| CAS No | 123-86-4 | 71-43-2 | 141-78-6 | 108-88-3 | 108-38-3 | 95-47-6 | 80-56-8 |
| Purity | 99.5% | 99.5% | 99.5% | 99.5% | 99% | 99% | 97% |
| Supplier | J&K Scientific ltd | J&K Scientific ltd | J&K Scientific ltd | Sinopharm Chemical Reagent Beijing Co., Ltd | J&K Scientific ltd | J&K Scientific ltd | J&K Scientific ltd |

In total, 24 binary mixtures of Ethyl acetate and the other six substances with identical concentrations of 10, 20, 50, and 100 mg m$^{-3}$, respectively, as well as 5 mixtures of all the seven substances with concentrations of 2, 5, 10, 20, and 50 mg m$^{-3}$, respectively, were prepared for this investigation.

## 2.3 Olfactometric measurements

The odour threshold concentration $C_{OT}$ (mg m$^{-3}$) of each pure substance, odour concentration and odour intensity ($C_{OD}^{olf}$ and $OI^{olf}$) were measured by dynamic olfactometry (AC'SCENT, USA) with *Forced-Choice Ascending Concentration Series Method*, which meet both (ASTM E679 - 04, 2011) and (EN13725, 2003) as shown in our previous study (Wu et al., 2015). The odour threshold concentration $C_{OT}$ is used to determine the specific odour mass $m_{OD}$ (mg ou$_E^{-1}$). The definition of $m_{OD}$ is analogue to the European reference odour mass (EROM) for n-Butanol in the EN 13725 (2003). The spreading of $m_{OD}$ in 1 m³ of pure air gives the unity odour concentration $C_{OD,0}$ = 1 ou$_E$ m$^{-3}$. The $m_{OD}$ can be calculated from the odour threshold concentration $C_{OT}$ and the unity odour concentration $C_{OD,0}$ = 1 ou$_E$ m$^{-3}$ by $m_{OD} = C_{OT} / C_{OD,0}$. $C_{OD}$ is then calculated by $C_{OD} = C / m_{OD}$ with the proper dimension of an odour concentration (ou$_E$ m$^{-3}$).

The relationship between $OI$ and $C_{OD}$ was measured with a suprathreshold gaseous substance by the olfactometer. Each panellist sniffed the three sample presentations from the olfactometer at each desired dilution level, one of which contains the gaseous substance while the other two are "blanks". The panellists continued to record $OI^{olf}$ of the presentation containing the dilute gaseous substance until $OI^{olf}$ reached the maximum of the intensity scale. $C_{OD}$ of the dilute gaseous substance at each of the dilution level was calculated as the ratio of chemical concentration to $m_{OD}$. Then the $OI$ - $C_{OD}$ relationship was derived by fitting the homologous $OI^{olf}$ and the $C_{OD}$ to the Weber-Fechner formula (Wu et al., 2015). To determine the intensity-concentration relationship, at least 5 different concentrations were offered to the panellists. The highest concentration was selected to reach an intensity of grade 4 (strong odour) to 5 (extremely strong).

The $C_{OD}^{olf}$ and the $OI^{olf}$ of the 29 mixtures of odouros substances were measured by dynamic olfactometry in the way as it was done with the pure substances. One mixture had to be eliminated because $C_{OD}^{olf}$ was below the detection limit of the olfactometer.

## 2.4 Data analysis

The relationship between the odour intensity $OI$ and the odour concentration $C_{OD}$ is described by the Weber-Fechner law for a certain substance $i$ according to

$$OI_i = k_i \ \log C_{OD,i} + 0.5$$

with the odour intensity $OI_i$, the logarithmically transformed odour concentration $C_{OD,i}$, and the slope $k_i$, which is often called Weber-Fechner coefficient. In general, the intercept $d$ of the Weber-Fechner law is determined by a regression analysis. However, for gaseous substances with $C_{OD} = 1$ ou$_E$ m$^{-3}$, the theoretical odour intensity ought to be $OI = 0.5$ according to the definition of the odour threshold concentration which states that 50% of the panellists perceive weak odour while the others perceive no odour. Hence, the Weber-Fechner law is adjusted to $OI = k \log C_{OD} + 0.5$, which means that the intercept of the linear relationship was fixed to $d = 0.5$ (Jiang et al., 2006; VDI 3882 Part 1, 1992). Even if the odour intensity scale is an ordinate scale, it is handled like a metric scale.

The model validation was done by a comparison of the empirically observed data, measured by olfactometry ($C^{olf}$ and $OI^{olf}$), and the modelled data for the four conversion methods using the root mean square error $RMSE$, the relative absolute error $RAE$ (Bennett et al., 2013), and the Nash-Sutcliffe model efficiency $NSE$ (Nash and Sutcliffe, 1970). The first two parameters describe the deviation of the model calculations from the empirical data. Therefore the ideal values of the $RMSE$ and the *RAE* are 0. *NSE* indicates the quality of how the model data fit to the line of identity with 1 as an optimal value. *NSE* < 0 indicates an even worse performance than using the mean value.

For the model selection, the models were compared by the use of the corrected Akaike´s information criterion *AIC$_c$* (Aho et al., 2014) which takes into account the number of parameters *P* by $K = P+1$. The aim of the *AIC$_c$* is to find the simplest model possible and prevent over-fitting by determining a relative ranking between models with the best showing the lowest value of *AIC$_c$*.

The statistical parameters are calculated according to

$$RMSE = \sqrt{\frac{1}{n}\sum_{i=1}^{n}(O_i - M_i)^2}$$

$$RAE = \sum_{i=1}^{n}|O_i - M_i| \Big/ \sum_{i=1}^{n}|O_i - \overline{O}_i|$$

$$NSE = 1 - \frac{\sum_{i=1}^{n}(O_i - M_i)^2}{\sum_{i=1}^{n}(O_i - \overline{O}_i)^2}$$

$$AIC_c = AIC + 2\frac{K+1}{N-K-1}$$

$$AIC = N \log RMSE^2 + 2K$$

with the empirically observed data $O_i$ and the modelled data $M_i$.

## 3. Results

### 3.1 Measurements of single substances

For the seven substances of Tab. 2, the odour threshold concentration $C_{OT}$ (mg m$^{-3}$), the derived specific odour mass $m_{OD}$ (mg ou$_E^{-1}$), and the slope *k* of the Weber-Fechner law were measured and are presented in Tab. 3. In addition, the range of expected $C_{OT}$ and $m_{OD}$ is given in the last line of the table, derived from the literature (Nagata, 1990; Schauberger et al., 2011). The range is very broad and, for m-Xylene and α-Pinene, even outside the measured value. This fact will be further discussed in Section 4.1.

Tab. 3: Olfactometrically measured odour threshold concentration $C_{OT}$ (mg m$^{-3}$), the derived specific odour mass $m_{OD}$ (mg ou$_E^{-1}$), and the slope $k$ (± standard deviation SD) of the Weber-Fechner law $OI = k \log C_{OD} + 0.5$ with the odour intensity $OI$ and the odour concentration $C_{OD}$ (ou$_E$ m$^{-3}$) for the seven odorous monomolecular substances. The expected values were derived from the literature.

| Substance | Butyl acetate | Benzene | Ethyl acetate | Toluene | m-Xylene | o-Xylene | α-Pinene |
|---|---|---|---|---|---|---|---|
| $C_{OT}$ and $m_{OD}$ | 1.337 | 8.624 | 2.680 | 3.365 | 1.610 | 3.284 | 4.182 |
| Slope $k$ (± SD) | 2.99 ± 0.10 | 2.59 ± 0.06 | 2.38 ± 0.04 | 2.98 ± 0.06 | 2.92 ± 0.013 | 3.19 ± 0.05 | 3.31 ± 0.17 |
| Correlation coefficient $r$ | 0.991 | 0.995 | 0.999 | 0.997 | 0.999 | 0.996 | 0.990 |
| $p$ value | < 0.001 | < 0.001 | < 0.001 | < 0.001 | < 0.001 | < 0.001 | < 0.001 |
| Expected $C_{OT}$ and $m_{OD}$ | 0.030 – 34.900[a] | 0.507 – 13.406[a] | 0.021 – 8.191[a] | 0.080 – 5.684[a] | 0.180 – 1.430[b] | 1.680 – 3.760[b] | 0.100 – 3.000[a] |

[a] Schauberger et al. (2011); [b] Yan et al. (2014b)

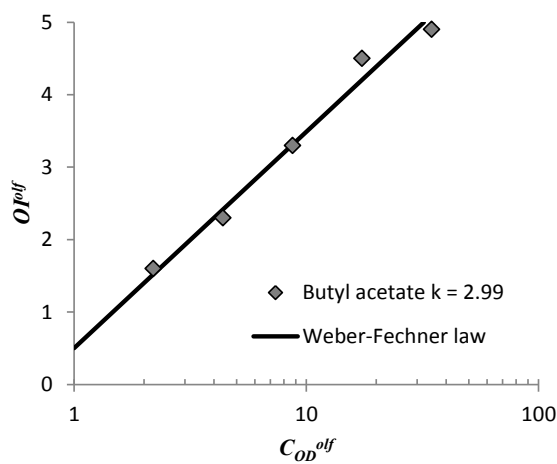

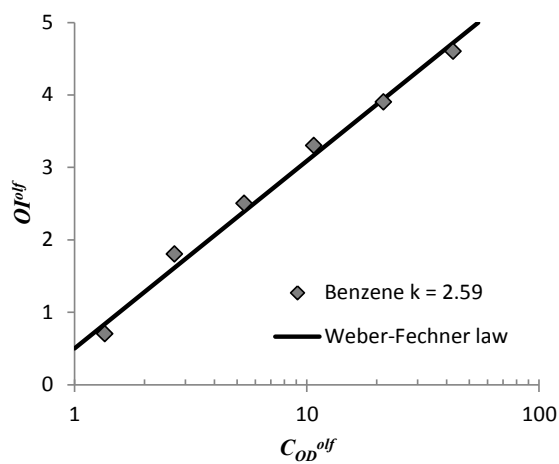

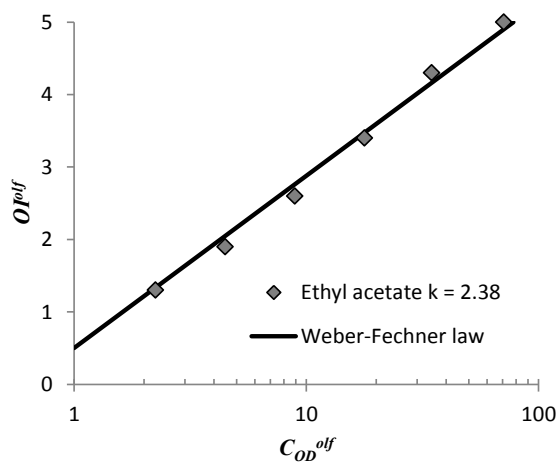

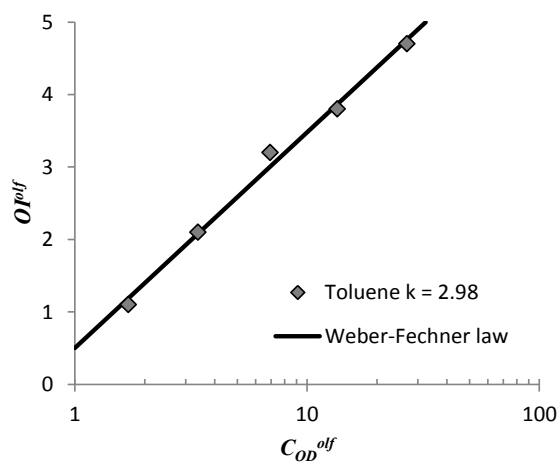

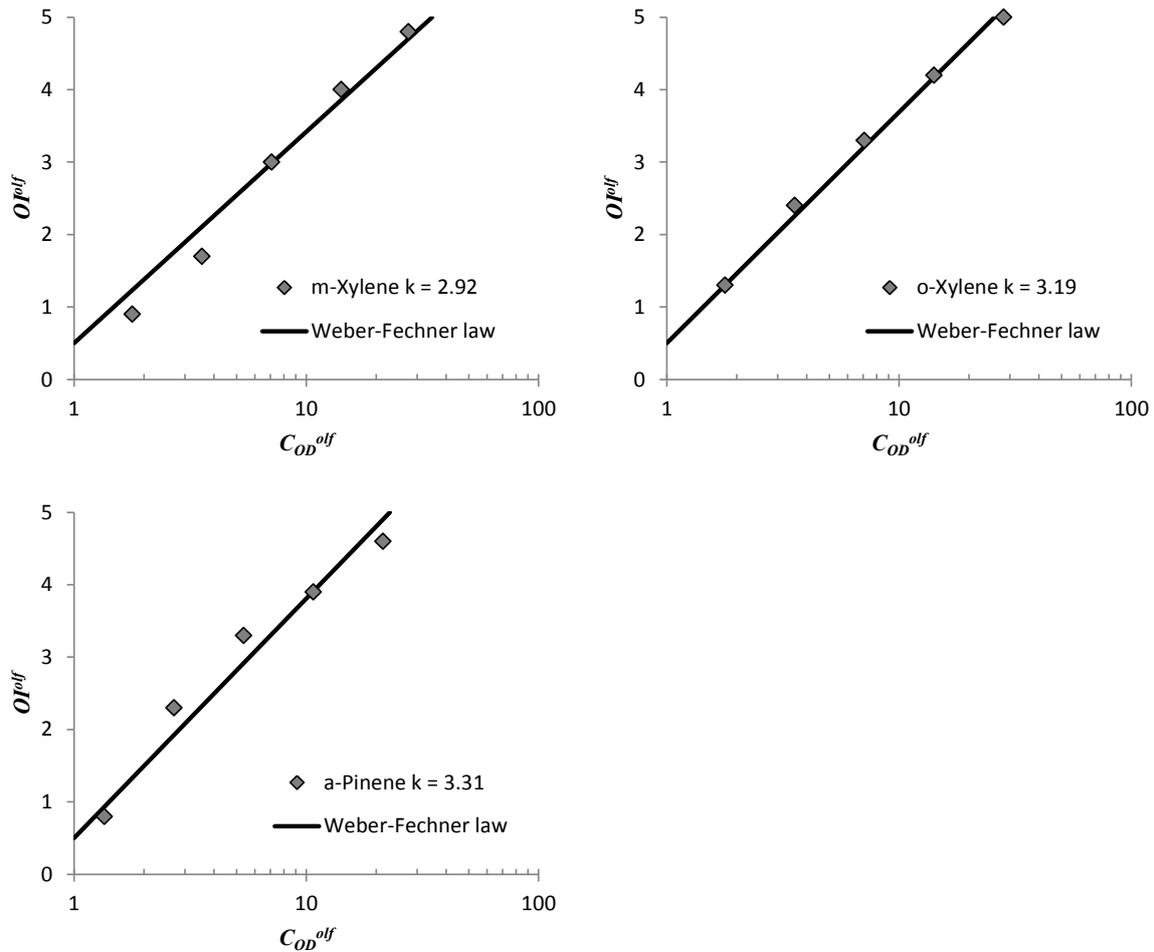

Fig. 2: Relationship between odour intensity $OI^{olf}$ and odour concentration $C_{OD}^{olf}$ (ou$_E$ m$^{-3}$) and the fitted Weber-Fechner law $OI = k \log C_{OD} + 0.5$ for the seven odorous monomolecular substances.

In Fig. 2, the relationship between the measured odour concentrations $C_{OD}^{olf}$ and the odour intensities $OI^{olf}$ for the single substances as well as the fitted Weber-Fechner law are shown. At least 5 data points were available for each linear regression. The regression coefficient $r$ and the $p$ values show a high level of significance.

The linear regression of the odour concentration and the odour intensity of all seven substances show a high correlation coefficient $r$, with a $p$ value below 0.1%. This gives the evidence that the Weber-Fechner law is a good assumption for this relationship. The slope of the Weber-Fechner law is in the range between 2.38 (Ethyl acetate) and 3.31 for α-Pinene. Ethyl acetate has the lowest slope and was selected as a reference substance.

### 3.2 Comparison of the conversion methods

To evaluate the four conversion methods outlined in Section 2.1, the calculated values were compared with the odour concentrations $C_{OD}^{olf}$ and the odour intensities $OI^{olf}$ measured by the olfactometer.

The perception of odour is not represented by the odour concentration $C_{OD}$ itself, because the perceived intensity is proportional to the logarithmically transformed stimulus. Therefore the crucial point of the conversion will be the agreement between $OI^{olf}$ and the four converted odour intensities $OI^C$ and $OI^{SOAV}$, $SOI$ and $OI^{EOC}$. The perfect conversion is represented by the line of identity. This agreement is tested by the *NSE* metrics.

The comparison of the calculated and the measured values is depicted in Fig. 3. The 23 binary mixtures and the 5 mixtures of all seven substances are presented in different symbols.

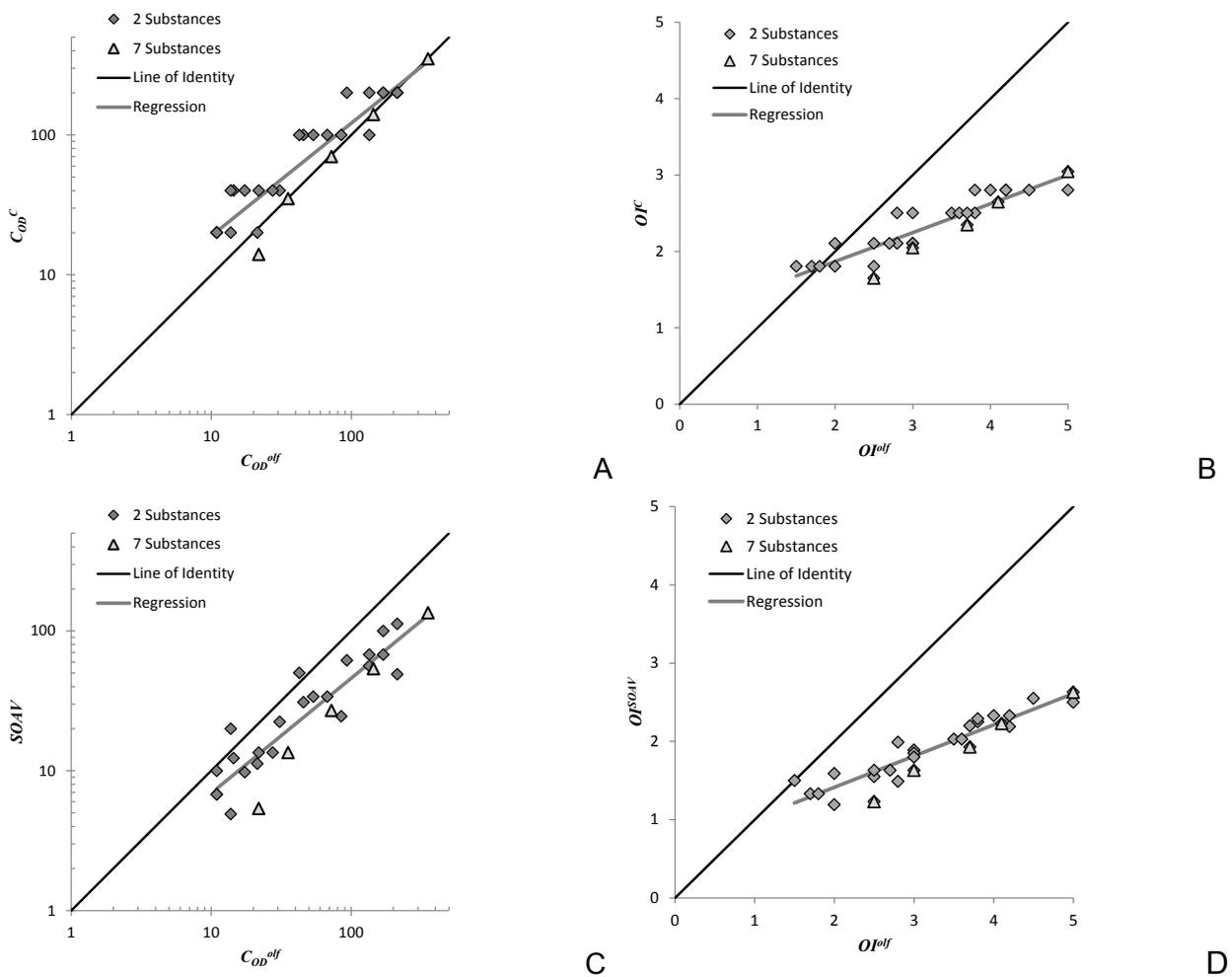

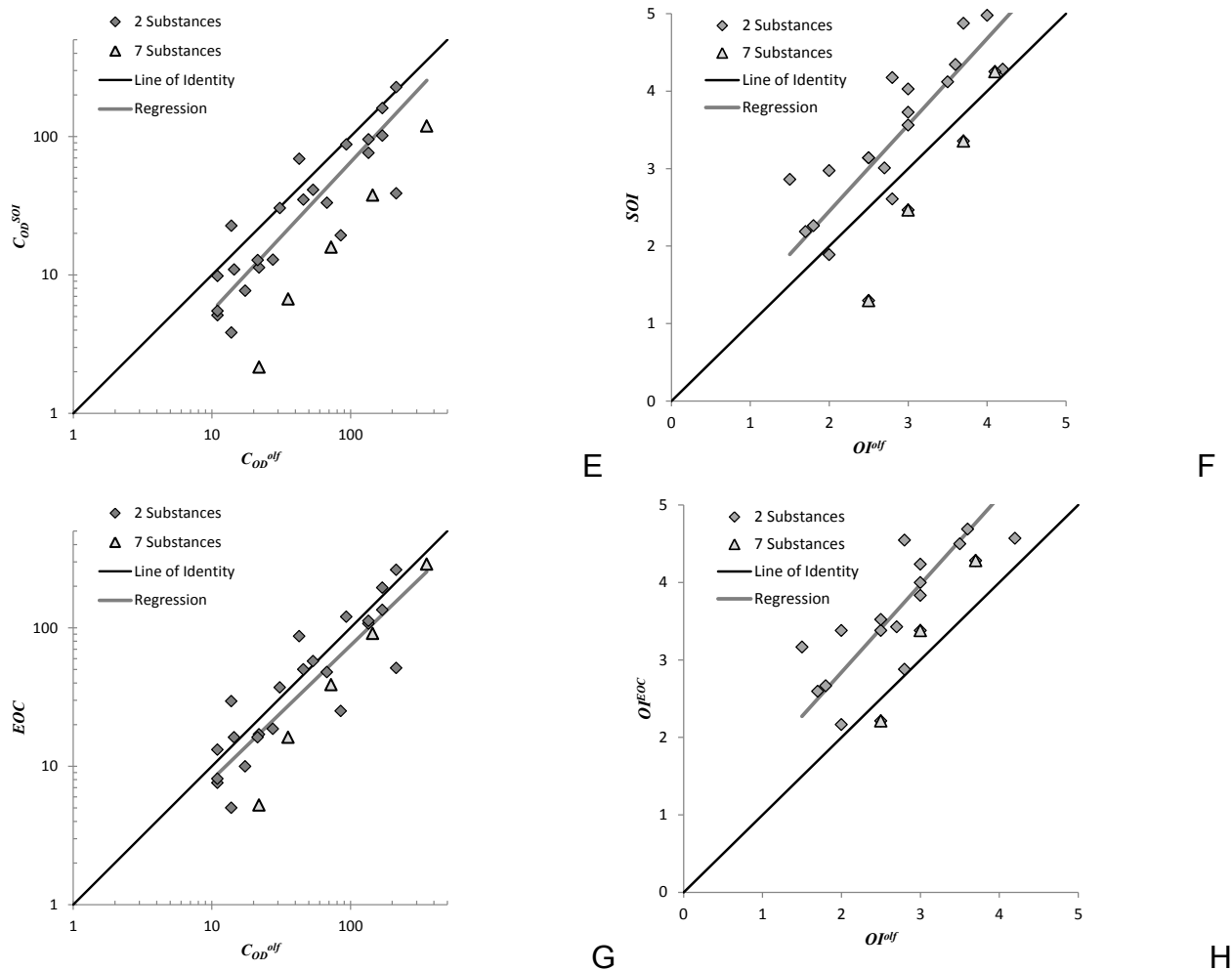

Fig. 3: Comparison of the converted odour concentrations $C_{OD}^{C}$ (A), $SOAV$ (C), $C_{OD}^{SOI}$, (E), and $EOC$ (G) ($ou_E$ m$^{-3}$) with $C_{OD}^{olf}$ ($ou_E$ m$^{-3}$) and the converted odour intensities $OI^{C}$ (B), $OI^{SOAV}$ (D), $SOI$ (F), and $OI^{EOC}$ (H) with the $OI^{olf}$. for the 23 binary mixtures and the 5 mixtures of all seven substances.

The general impression, when looking at Fig. 3, is that the results of the comparison improve with the increasing complexity of the conversion methods. This is somehow expected. The results for the comparison of concentrations are generally better than those for the odour intensities. The statistics of the conversion methods are given in Tab. 4.

The converted odour concentrations and odour intensities by the first two methods show the weakest quality (Fig. 3 A to D). For the first conversion method, the calculated odour concentrations for the 23 binary mixtures are over-estimated; those for the five mixtures of the seven substances are in line with the measurements (Fig. 3A). This conversion method will not provide odour intensities which are close to those measured by the olfactometer. Instead, the odour intensities are severely under-estimated.

The $SOAV$ needs the odour threshold concentration $C_{OT,i}$ for each substance to calculate the odour concentration for the entire mixture. The conversion shows a shallower slope compared to the ideal conversion and an underestimation for the investigated concentrations of about 43% of the odour concentrations (Fig. 3 C and D). The comparison of the odour intensities shows very similar results. For higher odour concentrations, the calculated odour intensities $OI^{SOAV}$ are too low, represented by the shallow slope. The reason for this is the use of a unity slope $k = 1$, because no further information is available for the conversion.

Tab. 4: Statistics of the conversion by the four methods. Linear regression of the logarithmically transformed odour concentration measured by the olfactometer $C_{OD}{}^{olf}$ (ou$_E$ m$^{-3}$) and the converted odour concentrations $C_{OD}{}^{C}$, $SOAV$, $C_{OD}{}^{SOI}$, and $EOC$ (ou$_E$ m$^{-3}$) as well as the odour intensity $OI^{olf}$ (-) from the olfactometer and the converted odour intensities $OI^{C}$, $OI^{SOAV}$, $SOI$, and $OI^{EOC}$ (-). The model evaluation was performed by the root mean square error $RMSE$, the relative absolute error $RAE$, the Nash-Sutcliffe model efficiency $NSE$, and the corrected Akaike's information criterion $AIC_c$. The values of the best performing models are highlighted in bold. The models which need further calibration by the olfactometric measurements are presented in grey.

| Conversion method | Linear regression | RMSE | RAE | NSE | $AIC_c$ |
|---|---|---|---|---|---|
| **Odour Concentration** | | | | | |
| $C$ | log $C_{OD}{}^{C}$ = 0.8055 log $C^{olf}$ + 0.4733 | 0.227 | 0.249 | 0.751 | -31.82 |
| $SOAV$ | log $SOAV$ = 0.8227 log $C^{olf}$ + 0.0164 | 0.335 | 0.543 | 0.457 | -20.27 |
| $SOI$ | log $C^{SOI}$ = 0.9471 log $C^{olf}$ - 0.2023 | 0.411 | 0.819 | 0.181 | -13.17 |
| $EOC$ | log $EOC$ = 0.9688 log $C^{olf}$ - 0.0651 | **0.263** | **0.334** | **0.666** | **-24.08** |
| **Odour Intensity** | | | | | |
| $C$ | $OI^{C}$ = 0.3778 $OI^{olf}$ + 1.114 | 1.08 | 1.28 | -0.28 | 6.02 |
| $SOAV$ | $OI^{SOAV}$ = 0.4001 $OI^{olf}$ + 0.6125 | 1.44 | 2.30 | -1.30 | 15.19 |
| $SOI$ | $SOI$ = 1.1153 $OI^{olf}$ + 0.2206 | **0.87** | **0.85** | **0.15** | **5.18** |
| $EOC$ | $OI^{EOC}$ = 1.1352 $OI^{olf}$ + 0.5696 | 1.13 | 1.41 | -0.41 | 11.36 |

The odour concentration, calculated by the $SOI$, shows a good correspondence with the line of identity with a slope of 0.9471. The converted odour concentration underestimates the measured odour concentration by about 37% (Fig. 3E). This under-estimation is even more pronounced for the mixtures of the seven substances. The regression line for the odour intensity shows a good agreement with the line of identity (Fig. 3F). The slope of the linear regression is 1.12 which results in an overestimation of about 0.2 grades for a high odour intensity of grade 5 (Tab. 4).

The equivalent odour concentration $EOC_{EA}$ shows a slope of 0.9688 which is close to the line of identity with a weak underestimation of about 13% (Fig. 3G). The regression line of the resulting odour intensities $OI^{EOC}$ lies parallel to the line of identity with a slope of 1.14 and an overestimation of about 0.6 grades of the 5 grade intensity scale (Fig. 3H).

The last two conversion methods, *SOI* and *EOC*, yield the best results. The regression lines for the odour intensity (Fig. 3F and Fig. 3H) show a good agreement with the line of identity. Therefore an additional calibration to adapt the slope to the line of identity is not needed.

The statistics of the four conversion methods are summarised in Tab. 4. The two models *SOI* and *EOC* are in the main focus of the evaluation, because they neeed no further calibration by olfactometric measurements, shown in black. The best results for the conversion of the odour concentrations are provided by the *EOC* method for all four parameters. Especially the parameter *NSE* which evaluates the fit with the line of identity is distinctly higher for the *EOC* than for the *SOI*. For the odour intensity, the statistic parameters show the best results for the *SOI* method. In both cases the respective method for a certain model goal shows the best results. For the odour intensity, the *SOI* conversion method, for the odour concentration, the *EOC* method performs best.

## 4. Discussion

The assessment of the odour concentration and/or the odour intensity of a mixture of odorous substances is not trivial, but a series of methods exist, as outlined in Section 2.1. A major motivation is the substitution of direct (sensory, olfactometric) methods to measure the odour concentration which are expensive and time-consuming by indirect methods, e.g. concentration measurements of (mixtures of) odorous substances. The measurement of the concentration of such substances is well established and is used on an operational basis by national environmental agencies.

A common method for the measurement of odorous substances is the fully automatic gas chromatography with several detectors like mass spectroscopy or flame ionization detector with a sampling cycle of about 30 min due to the enrichment of the odorous substances in adsorption tubes (e.g. Niemeyer et al. (2010)). A higher temporal resolution and a lower detection threshold can be performed by proton transfer reaction PTR (e.g. Feilberg et al. (2010b)) or the selected ion flow tube mass spectrometry (SIFT-MS) (e.g. Heynderickx et al. (2013)). All these measuring methodologies provide the concentration of chemical substances. For each of these substances, if odorous, the related characteristics for the perception of a smell (odour threshold concentration and intensity) have to be known.

## 4.1 Measurements of single substances

The investigation presented here was performed for seven substances (Butyl acetate, Benzene, Ethyl acetate, Toluene, m-Xylene, o-Xylene and α-Pinene), which are emitted by petrochemical plants (Muezzinoglu and Dincer, 2005) or chemical waste treatment plants (Schauberger et al., 2011). Some of these substances are also emitted by municipal solid waste landfills (Saral et al., 2009; Wenjing et al., 2015). For these substances, the odour threshold concentration $C_{OT}$ and the slope of the Weber Fechner law were olfactometrically measured and compared with values derived from a literature review (Schauberger et al., 2011). We couldn't find any evidence for the high uncertainty of these values shown by the broad range (e.g. for the seven substances in Tab. 3). Due to the age of some investigations, some of these values could be based on improper measurement techniques. The odour threshold concentration $C_{OT}$ is available for many chemical substances, but only a few investigations were performed to determine the coefficient of the Weber-Fechner law (van Ruth and O'Connor, 2001; Wu et al., 2015; Yan et al., 2014a). The parameters for 12 odorous substances which are used for the Korean air quality assurance program are summarised by Kim and Park (2008), based on the measurements by Nagata (2003). These parameters are given as mixing ratios (ppm) instead of densities (mg m$^{-3}$).

The unknown accuracy of $C_{OT}$ found in literature has long been a constraint to the application and communication of research on odour conversion methods, which need the $C_{OT}$ value (Capelli et al., 2008; Feilberg et al., 2010a). In this study, $C_{OT}$ and $C_{OD}$ were determined by the same group of well-trained sniffing panellists to minimize the intra-laboratory error. Furthermore, a dynamic olfactometry and standard method which meet both ASTM E679-04 and EN 13725 standard was used to carry out the olfactometric experiment, so as to reduce the inter-laboratory error.

The relationship between odour intensity $OI$ and odour concentration $C_{OD}$ is described by several functions like the Weber-Fechner law (exponential function) or the Stevens law (power function) (Bundy et al., 1997; Guo et al., 2001; Sarkar and Hobbs, 2002; Sarkar et al., 2003; Zhang et al., 2002). The most frequently used model is the Weber-Fechner law, showing a good fit to empirical data (Nicolai et al., 2000; Sarkar and Hobbs, 2002). The intensity scale can vary between 5 and 12 grades (Yu et al., 2010), even a scale with 13 grades is in use (Reinbach et al., 2011). Yu et al. (2010) give an overview of the used scales to determine odour intensity. Guo et al. (2006) show the problems in the translation of various grade scales. A comparison of several scales and the concentrations of n-Butanol for the various grades of the scales are summarised by McGinley and McGinley (2000). Especially the strongest intensity (last grade; very strong or extremely unbearable odour) shows a wide variety of n-Butanol concentrations (between 20480 ppm for the grade 12 and 1550 ppm for grade 8).

In general the Weber-Fechner law is a two-parametric function with the slope $k$ and the intercept $d$. Jiang et al. (2006) and VDI 3882 Part 1 (1992) suggest a fixed value of $d_i$ = 0.5 for the intercept due to the fact that only 50% of the panellists perceive odour for an odour concentration $C_{OD,i}$ = 1 ou$_E$ m$^{-3}$. Yu et al. (2010) summarised the Weber-Fechner law of several authors for livestock odour. If those data are eliminated which are not based on the Weber-Fechner law, then the mean intercept gives $d$ = 0.66, which is close to the proposed constant value of $d$ = 0.5. The slope gives then $k$ = 1.96.

## 4.2 Comparison of the conversion methods

The conversion from the concentration of single substances to the odour concentrations and odour intensities of the odorous mixture is demonstrated by four methods with increasing complexity, depending on the necessary prerequisites for the calculation (Tab. 1).

The first method uses the sum of the concentrations as a surrogate for the odour concentration and odour intensity of the mixture $C_{OD}^C$ and $OI^C$, respectively. This method cannot be used without additional olfactometric measurements of the relationship between odour intensity and odour concentration by an olfactometer to determine the slope of the regression line (Fig. 3A and B). This can be seen by the deviation of the regression line of the odour intensity from the line of identity. Nevertheless, this method is used for single substances (e.g. H$_2$S (Dincer and Muezzinoglu, 2007; Gostelow and Parsons, 2000; Gostelow et al., 2001)) or a group of substances (e.g. entire concentration of VOCs (Capelli et al., 2013)).

The second concept which includes the odour threshold concentration as a physiological relevant value is the odour activity value $OAV$ concept and its sum for several substances $SOAV$ (Parker et al., 2012). Even if the $SOAV$ is a dimensionless number, it is often interpreted as an odour concentration (e.g. Wenjing et al. (2015), Capelli et al. (2008)), called theoretical odour concentration. Therefore we suggest to use the specific odour mass of an individual substance $m_{OD,i}$, which is based on the odour threshold concentration $C_{OT,i}$, to yield a proper dimension for an odour concentration. One of the advantages of the $SOAV$ is the assessment of the relative importance of the perception of a single substance of a mixture (Laor et al., 2014; Lee et al., 2013). The relative importance of each individual substance can be calculated by the portion $P_i$ of the individual compound $P_i = OAV_i/SOAV$. The portion $P_i$ and the individual $OAV_i$ are often used to identify the contribution of the individual substances to the entire odour pollution. One of the uncertainties of the calculation of the $OAV$ is based on the imprecision of reliable odour threshold concentrations (Capelli et al., 2013; Schauberger et al., 2011; Wenjing et al., 2015). In a majority of previous studies, odour threshold concentrations from disparate literature sources are cited to calculate $OAV$. However, the odour threshold concentration of a certain substance may differ over several orders of magnitude due to the discrepancy of sniffing panellists and olfactory methods. In order to minimize this imprecision, we measured odour threshold concentrations of the seven pure substances as well as the odour concentration and the odour intensity of the mixtures by the same panellists. In several studies the calculated $OAV$ was compared with the odour concentration measured by an olfactometer. Capelli et al. (2013) reported such a comparison for environmental odour emitted by an industrial complex including steel industry, different chemical industries for the production of polypropylene and for treatment plants for waste waters and solid waste. By the use of the $OAV$ instead of the concentrations of the VOCs, the coefficient of determination could be increased from $r^2 = 0.393$ to $r^2 = 0.836$. Blazy et al. (2015) found a coefficient of determination of $r^2 = 0.87$ for odour emitted by a pig slaughterhouse sludge composting and storage plant. For livestock odour (pigs and dairy) Parker et al. (2012) compared the $OAV$ with olfactometric measurements of the odour concentration. The coefficient of determination was in the range between $0.16 < r^2 < 0.52$. By a multiple linear regression of the individual $OAV_i$ the model fit could be increased to $0.62 < r^2 < 0.96$, depending on the number of odorous substances included in the regression model. Even if the model fit seems very high, the decision which substance to be included as well as the choice of the coefficients of the multiple linear model has to be done on an individual basis. In addition, the weakness of this approach lies in the fact that the coefficients are variable as the number of odorous substances included in the regression model is changing, and occasionally occurring negative regression coefficients can't be interpreted properly. This means that this approach cannot be applied without olfactometric measurements; therefore it cannot be used universally (Akdeniz et al., 2012).

The $SOI$ conversion method of Kim and Park (2008) uses a relationship between the concentration $C_i$ and the related odour intensity $OI_i$ with a modified Weber-Fechner law by $OI_i = k_i \log C + d_{C,i}$ based on (Nagata, 2003). In this version of the Weber-Fechner law, the intercept $d_{C,i}$ includes not only the odour threshold concentration $C_{OT,i}$ but also the intercept $d_i$ of the Weber-Fechner law.

Lee et al. (2013), for field measurements, could show that the correlation between the $OAV$ and the odour intensity by using the $SOI$ could be increased from $r^2 = 0.216$ to $r^2 = 0.518$.

The equivalent odour concentration $EOC$ for an odorous mixture is a conversion method introduced here for the first time. The difference between the $SOI$ and the $EOC$ conversion is the order of calculations. For the $SOI$ method, the individual odour intensity $OI_i$ is calculated directly from the concentration of the selected substance (Kim and Park, 2008). Afterwards the odour concentration of the mixture is calculated for the reference substance, in our case for Ethyl acetate. The $EOC$ method starts with the calculation of the odour concentration for a certain substance and calculates the odour intensity of this substance afterwards. To compare the two conversion methods $SOI$ and $EOC$, we used the same slope of the Weber-Fechner law (Tab. 3).

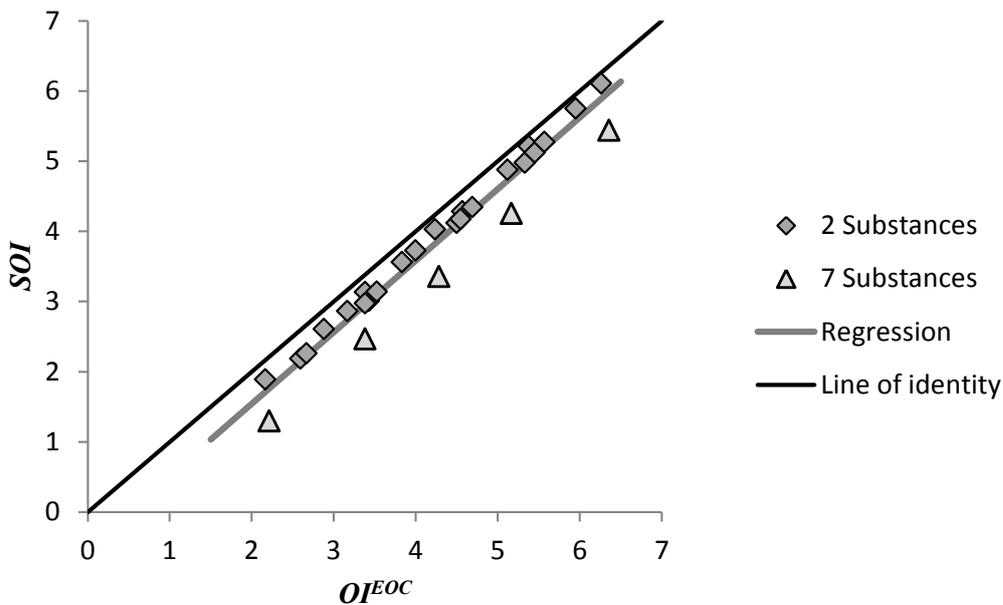

Fig. 4: Comparison of the odour intensity, calculated by the *SOI* and the *EOC* conversion methods as a scatter gram for the 23 binary mixtures and the 5 mixtures of all seven substances.

A comparison of the odour intensity, calculated by the *SOI* and the *EOC* conversion methods, is shown in Fig. 4. The linear regression is close to the line of identity which means that the two methods show only small deviations from each other. The good fit of the converted odour intensity with the olfactometric measurements for the selected odour mixtures shows that no additional olfactometric measurements are necessary to calibrate the calculations with the measurements. Therefore it can be assumed that the goal of an objective conversion is reached.

The evaluation of the conversion methods was done by a limited number of seven odorous substances and the mixtures of these substances. The substances were selected to apply this approach to a case study for a thermal waste treatment plant published by Schauberger et al. (2011). It is admitted that these selected substances are chemically similar to each other, and any particular selection can have a considerable influence on the outcome. To get a broader empirical basis for the model evaluation, the conversion methods should be compared to other available datasets. Kim (2011), for example, uses a mixture of the following odorous substances: hydrogen sulphide, and five aldehydes (Acetaldehyde, Propionaldehyde, Butyraldehyde, *iso*-Valeraldehyde, and Valeraldehyde). Besides the former study, the investigation by Capelli et al. (2012) could be used in the same way for typical industrial odour emissions.

All the discussed concepts are based on the working hypothesis that the mixture of odorous substances behaves additively, which is only a rough estimate (Thomas-Danguin et al., 2014). This means that no interactions between the substances take place. For environmental odour, usually several effects are observed. Grosch (2001) could show that odorants with a higher odour activity value of a single compound $OAV_i$, (or individual odour concentration $C_{OD,i}$) are frequently essential for the aroma. However, there are exceptions where odorants with high OAVs are suppressed in the aroma and compounds with lower OAVs are important contributors. The effect of mixing of multiple odorants is described by Kim and Kim (2014a) for food odour showing that augmentation, as well as protective effects, and other effects like masking or dominance by a stronger component, and synergistic effects can be expected. Especially for odorous mixtures of pure odorant and mixed olfactory/trigeminal stimulus, the concepts of conversion have to show proof for the applicability. These uncertainties highlight that the use of the chemical concentrations should be limited to those cases where the odour concentration is - for some reasons - not directly measurable.

## 5. Conclusions

The investigation presented here successfully shows ways to convert the concentrations of single substances to odour concentrations and odour intensities of an odorous mixture. For frequently occurring chemical substances, four conversion concepts with increasing complexity were compared. It could be shown that only those conversion methods deliver reliable values which use not only the odour threshold concentration but also the slope of the Weber-Fechner law to include the sensitivity of the odour perception of the individual substances. These two methods (*SOI* and *EOC*, see Fig. 2 and Tab. 4) fulfil the criteria of an objective conversion without the need of a further calibration by additional olfactometric measurements.

The uncertainty of the odour threshold concentration $C_{OT}$ in the literature is a weak point for all three methods of conversion which need this value. Therefore the values for frequently needed substances should be determined in a joint endeavour to get comparable results.

The determination of odour concentrations and odour intensities of odorous mixtures from the concentrations of individual substances has its merits as, if successful as in the case of the *SOI* and *EOC* conversion methods, olfactometric measurements can be avoided. Such methods allow now to measure time series of odour emission rates, ambient odour concentrations as well as the back-calculation of emission flow rates by inverse modelling.

## 6. Acknowledgements

This work was jointly supported by the National Natural Science Foundation of China (Nos. 21277011, 21407008 and 21576023), the major accident prevention key projects of State Administration of Work Safety (No. Beijing-0003-2015AQ). The cooperation between China and Austria was funded by the Eurasia-Pacific Uninet (No. 18/2014).

# 7. Nomenclature

| | |
|---|---|
| $C$ | Concentration of the chemical substance (mg m$^{-3}$) |
| $C_{OT}$ | Odour threshold concentration (mg m$^{-3}$) |
| $m_{OD}$ | Specific odour mass (mg ou$_E^{-3}$) |
| $C_{OD,0}$ | Unity odour concentration ($C_{OD,0}$ = 1 ou$_E$ m$^{-3}$) |
| $C_{OD}$ | Odour concentration (ou$_E$ m$^{-1}$) |
| $OI$ | Odour intensity (-) |
| $C_{OD}^{olf}$ | Odour concentration measured by an olfactometer |
| $OI^{olf}$ | Odour intensity measured by an olfactometer |
| $k$ | Slope of the Weber-Fechner law $OI = k \log C_{OD} + d$ |
| $k_c$ | Proportional constant of the $C$ method     $d$     Intercept of the Weber-Fechner, with $d$ = 0.5 |
| $C_{OD}^{C}$ | Odour concentration, converted by the concentration of the chemical substance (ou$_E$ m$^{-3}$) |
| $OI^C$ | Odour intensity, converted by the concentration of the chemical substance (-) |
| $OAV$ | Odour activity value (ou$_E$ m$^{-3}$) |
| $SOAV$ | Sum of the odour activity value (ou$_E$ m$^{-3}$) |
| $OI^{SOAV}$ | Odour intensity, converted by the sum of the odour activity value (-) |
| $C_{OD}^{SOI}$ | Odour concentration, converted by the sum of the odour intensity (ou$_E$ m$^{-3}$) |
| $SOI$ | Sum of the odour intensity (-) |
| $EOC$ | Equivalent odour concentration (ou$_E$ m$^{-3}$) |
| $OI^{EOC}$ | Odour intensity, converted by the equivalent odour concentration (-) |